\documentclass{article} 
\usepackage{lscape,a4wide,amssymb,amsmath,cite,epsfig,psfrag}

\def \colthree {\mbox{$\bar{\mathbf{3}}_c$}}
\def \colsix {\mbox{$\mathbf{6}_c$}}
\def \FSthreeone {\mbox{$\bar{\mathbf{3}}_1$}}
\def \FSsixone {\mbox{$\mathbf{6}_1$}}
\def \FSthreezero {\mbox{$\bar{\mathbf{3}}_0$}}
\def \FSsixzero {\mbox{$\mathbf{6}_0$}}

\newcommand{\SpSp}[2]{ \mbox{$\vec{S}_{#1}.\vec{S}_{#2}$}}
\newcommand{\lala}[2]{ \mbox{$\frac{\vec{\lambda}_{#1}}{2} \cdot
    \frac{\vec{\lambda}_{#2}}{2}$}}

\newcommand{\delij}[2]{ \mbox{$\delta(\vec{r}_{#1 #2})$}} 

\begin{document} 
 
\begin{titlepage} 
 
  \baselineskip=18pt \vskip 0.9in
\begin{center} 
  {\bf \Large A model realisation of the Jaffe-Wilczek correlation for pentaquarks}\\
  \vspace*{0.3in}
  {\large J.J.Dudek}\footnote{\tt{e-mail: dudek@thphys.ox.ac.uk}}\\
  \vspace{.1in} {\it Department of Physics - Theoretical Physics, University of Oxford,\\
    1 Keble Rd., Oxford OX1 3NP, UK} \\
  \vspace{0.1in}
 
\end{center} 
 
\begin{abstract} 
  
We discuss a realisation of the pentaquark structure proposed by Jaffe and Wilczek within a simple
quark model with colour-spin contact interactions and coloured
harmonic confinement, which accurately describes the $\Delta-N$ splitting.  In this model spatially compact diquarks are formed in the pentaquark but no such compact object exists in the nucleon. The colour-spin attraction
 brings the Jaffe-Wilczek-like state down to a low mass, compatible with the experimental observation and below that of the naive ground state with all $S$-waves. We find, however, that although these trends are maintained,
the extreme effects observed do not survive the required ``smearing''
of the delta function contact interaction. We also demonstrate the weakness of the ``schematic'' approximation when applied to a system containing a $P$-wave. An estimate of the anti-charmed pentaquark mass is made which is in line with the Jaffe-Wilczek prediction and  significantly less than the value reported by the H1 collaboration.
 
\end{abstract} 

\end{titlepage} 

\section{Introduction}

In the wake of the possible discovery of baryon with positive
strangeness (the $\Theta$)\cite{thetaexp}, phenomenological models have arisen
designed to explain its low mass ($\sim 1540$MeV) and narrow width ($<
10$MeV). For an early review see \cite{fechad03}.

One such model is due to Jaffe and Wilczek\cite{jw}. They propose diquark
correlations of low mass in a relative $P$-wave, giving the state
overall positive parity in line with the Chiral Soliton Model
prediction\cite{dpp}. 

The Jaffe-Wilczek model has two diquarks each of spin-0, flavour
$\bar{\mathbf{3}}_F$ and colour $\bar{\mathbf{3}}_c$, which by Bose
statistics must couple symmetrically. To give an overall
$\overline{\mathbf{10}}_F$ the diquarks must couple to a
$\bar{\mathbf{6}}_F$, i.e. symmetrically. An overall colour singlet
requires antisymmetric diquark coupling to a $\mathbf{3}_c$ and hence
to be overall symmetric we are forced to introduce a $P$-wave between
the diquarks. A $P$-wave appears in many quark model treatments of
pentaquark structure, but, unlike in the Jaffe-Wilczek model, it is usually rather {\em ad
  hoc}.

This model has attracted
much attention, 
and is particularly appealing as the same diquark
correlations can be used to explain the enigmatic scalar mesons below
1GeV as being diquark-antidiquark states\cite{jaffescalar, FecTorn}. 

Our point of departure is to discuss the dynamics which gives rise to diquarks. This an open question which must eventually be answered directly from QCD, but we will show that such correlations
can appear even within simple quark potential models of the type often
used to describe the conventional hadron spectrum.

\section{Schematic Approximation, Quark Potential Model and a Jaffe-Wilczek-like state}

Previous investigations into pentaquark structure have often made use of the ``Schematic'' approximation to colour-spin forces\cite{schematic}. This approximation
discards spatial dependence, having an interaction potential
$V^\sigma_{\mathrm{SCH}} = - C \sum_{i,j} \lala{i}{j} \SpSp{i}{j}$,
where $C$ is a constant. For example in the total spin-0 channel this would give,
\begin{equation}\label{Schem}
    \langle \colthree,  \colthree | V^\sigma_{\mathrm{SCH}}| \colthree, \colthree
  \rangle = \begin{array}{c} -C \\ 0 \end{array} \left\{\begin{array}{l} 0 \otimes 0 \to 0 \\ 1 \otimes 1 \to 0 \end{array} \right. ,
\end{equation}
where e.g. $1 \otimes 1 \to 0$ indicates that 1 \& 2 and 3 \& 4 are each
coupled to spin-1 and then the two pairs coupled to total spin-0.

We will go beyond this approximation by allowing there to be a
non-trivial spatial dependence. In particular we shall see that the
schematic approximation, while capturing all the physics for spatially
symmetric states, is not sufficiently versatile to accurately describe
states containing a $P$-wave. This will be demonstrated by computing
the equivalent to \eqref{Schem} in a model with non-trivial spatial
dependence.

We now introduce the model 
which, in essence, is a standard quark potential model suitable for describing the light baryon and meson spectrum.

As a binding potential between quarks we take the coloured harmonic oscillator,
\begin{equation}\label{vqq}
  V(q_i q_j) = - a  \lala{i}{j}  \,(\vec{r}_i - \vec{r}_j)^2.
\end{equation}
This is chosen mainly for ease of calculation but has been used in the past
as an approximation to the more phenomenologically justified coloured
Coulomb + linear potential.

In addition to this we introduce a colour-spin contact interaction,
\begin{equation}\label{vcsqq}
  V^\sigma(q_i q_j) = -h \lala{i}{j} \, \SpSp{i}{j} \, \delij{i}{j}.
\end{equation}
It is this potential that will act to bind two quarks tightly into an
effective light diquark.

The fundamental mechanism binding the diquarks in the
Jaffe-Wilczek picture is unlikely to be anything so simple as a colour-spin contact
interaction, and indeed it may not be possible to consider diquarks in terms of constituent quarks with non-relativistic interactions at all. Diquarks may appear as degrees-of-freedom in the same way as constituent quarks do during chiral symmetry breaking. Despite this, colour-spin contact interactions have a considerable
phenomenological pedigree and may well be able to simulate quite
effectively the true QCD interaction responsible for the ``fine''
structure in hadrons and as such we will investigate their effect on the pentaquark system. The motivation we have for a delta-function interaction
is that of \cite{dgg}, the Breit-Fermi reduction of one-gluon exchange
between quarks, but delta-function interactions have also been used to model
instanton effects in light hadrons (see for example \cite{Brau} and references therein). 

\begin{figure} 
\begin{center} 
  \psfragscanon \psfrag{Q1}[]{$\mathbf{q_1}$}
  \psfrag{Q2}[]{$\mathbf{q_2}$} \psfrag{Q3}[]{$\mathbf{q_3}$}
  \psfrag{Q4}[l]{$\mathbf{q_4}$} \psfrag{Qbar}[]{$\bar{\mathbf{q}}$}
  \psfrag{T}[b]{$\vec{\tau}$} \psfrag{Rho}[]{$\sqrt{2}\vec{\rho}$}
  \psfrag{Nu}[l]{$\sqrt{2}\vec{\nu}$}
  \psfrag{Lam}[br]{$\vec{\lambda}$} \psfrag{Cm}[]{c.m.}
  
  \includegraphics[width=3in]{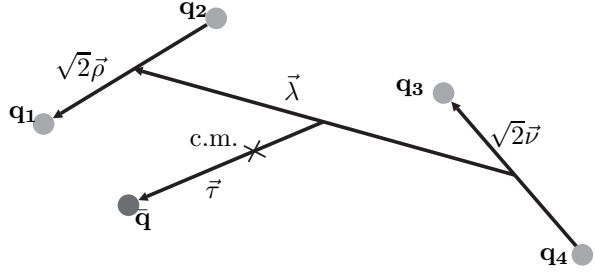}
\caption{Five-body co-ordinate set in the $q^4\bar{q}$ rest frame. \label{coords}}  
\end{center}
\end{figure}

Working in the $q^4\bar{q}$ rest frame, we define the internal variables,
$\vec{\tau},\vec{\lambda},\vec{\rho},\vec{\nu}$, by
\begin{eqnarray}\label{vars}
  \vec{r}_1 &=&  -\frac{\mu}{4m} \vec{\tau} + \frac{1}{2}
  \vec{\lambda} + \frac{1}{\sqrt{2}} \vec{\rho} \nonumber \\
  \vec{r}_2 &=&  -\frac{\mu}{4m} \vec{\tau} + \frac{1}{2}
  \vec{\lambda} - \frac{1}{\sqrt{2}} \vec{\rho} \nonumber \\
  \vec{r}_3 &=&  -\frac{\mu}{4m} \vec{\tau} - \frac{1}{2}
  \vec{\lambda} + \frac{1}{\sqrt{2}} \vec{\nu} \nonumber \\
  \vec{r}_4 &=&  -\frac{\mu}{4m} \vec{\tau} - \frac{1}{2}
  \vec{\lambda} - \frac{1}{\sqrt{2}} \vec{\nu} \nonumber \\
  \vec{r}_{\bar{q}} &=&  \frac{\mu}{m_{\bar{q}}}\vec{\tau},
\end{eqnarray}
which are clearly well suited to describing a Jaffe-Wilczek-like
configuration if the diquarks are spatially correlated (and see figure \ref{coords}). With these variables the kinetic energy
is\footnote{ $\mu$ is the $q^4, \bar{q}$ reduced mass, $\frac{m_{\bar{q}} 4m}{4m + m_{\bar{q}}}$.} 
\begin{equation}\label{kin}
  T(q^4\bar{q}) = \frac{\vec{\nabla}_{\tau}^2}{2 \mu}+
  \frac{\vec{\nabla}_{\lambda}^2}{2m}+\frac{\vec{\nabla}_{\rho}^2}{2m}+\frac{\vec{\nabla}_{\nu}^2}{2m},
\end{equation}
and we can write the Hamiltonian $H(q^4\bar{q}) = T(q^4\bar{q}) + V(q^4)+V(\bar{q}) +V^\sigma(q^4)+V^\sigma(\bar{q})$.

We can approximately solve for an eigenstate of this Hamiltonian using
the variational method. First introduce a trial wavefunction containing a
number of variational parameters and minimize the expectation value of
the Hamiltonian with respect to them. The trial
wavefunction has spatial form
\begin{equation}\label{psijw}
  \psi_m = N [\beta \lambda Y_{1m}(\hat \lambda) e^{- \beta^2
    \lambda^2 /2}][ Y_{00}(\hat \rho)e^{- \gamma^2 \rho^2 /2}][
  Y_{00}(\hat \nu) e^{- \gamma^2 \nu^2 /2}][ Y_{00}(\hat \tau) e^{-
    \alpha^2 \tau^2 /2}],
\end{equation}
which is a $P$-wave in the $\vec{\lambda}$ variable and an $S$-wave in
the others in the harmonic oscillator approximation. The
flavour-spin-colour structure is chosen to be $|\FSthreezero,
\FSthreezero\rangle_{\bar{\mathbf{6}},0} \otimes |\colthree,
\colthree\rangle_{\mathbf{3}_c}$, where this is a shorthand for quarks
1 \& 2 coupled to a $\bar{\mathbf{3}}$ of flavour, spin 0 and colour
$\bar{\mathbf{3}}$; quarks 3 \& 4 are coupled identically to 1 \& 2 and
then the pairs \{12\},\{34\} coupled to a  $\bar{\mathbf{6}}$ of
flavour, spin 0 and a $\mathbf{3}$ of colour. This is the
flavour-spin-colour structure of the Jaffe-Wilczek correlation. The antiquark couples trivially to give an overall colour singlet.

As stated this wavefunction (flavour-spin-colour-spatial) is only antisymmetric under exchange of
labels $1 \leftrightarrow 2$ or $3 \leftrightarrow 4$, but not for
example $1 \leftrightarrow 3$. This would seem to violate the
generalized Pauli principle. In fact we need not go to the trouble of 
antisymmetrising two particles if their wavepackets have very limited
overlap - we don't have to worry about electrons on the moon when we
solve the Schr\"{o}dinger equation for a hydrogen atom on Earth nor, more pertinently, do
we usually antisymmetrise the quarks in different nucleons when we study
the deuteron. Here we are proposing that the dynamics in the Hamiltonian
given earlier will be such that the wavefunctions of quarks in different
diquarks have very limited overlap such that we don't have to
antisymmetrise them. We will show that this assumption is consistent
when we variationally solve the Hamiltonian with this trial
wavefunction.

With the $|\colthree, \colthree\rangle_{\mathbf{3}_c}$ colour state
the expectation values of the potentials are
\begin{eqnarray}\label{Vjw}
  \langle \colthree,  \colthree | V(q^4)| \colthree, \colthree
  \rangle &=& \frac{5}{3} a (\rho^2 + \nu^2) + \frac{2}{3} a \lambda^2
  \nonumber \\
  \langle \colthree,  \colthree | V(\bar{q})| \colthree, \colthree
  \rangle &=& \frac{1}{3} a (\rho^2 + \nu^2) + \frac{1}{3} a \lambda^2+ \frac{4}{3} a \tau^2
  \nonumber \\
  \langle \colthree,  \colthree | V^\sigma(q^4)| \colthree, \colthree
  \rangle &=& \frac{2}{3} h \big(\SpSp{1}{2}  \delij{1}{2} + \SpSp{3}{4}
  \delij{3}{4} \big) \nonumber \\
&+& \frac{1}{6}h \big( \SpSp{1}{3}  \delij{1}{3} +\SpSp{1}{4}
\delij{1}{4} +\SpSp{2}{3}  \delij{2}{3} +\SpSp{2}{4}
\delij{2}{4}\big) \nonumber \\
  \langle \colthree,  \colthree | V^\sigma(\bar{q})| \colthree, \colthree
  \rangle &=& \frac{1}{3} h \, \vec{S}(q^4).\vec{S}(\bar{q}) \delij{1}{\bar{q}}.
\end{eqnarray}
With the spatial wavefunction \eqref{psijw} we have
\begin{eqnarray}\label{exp}
  \langle \delij{1}{2} \rangle = \langle \delij{3}{4} \rangle =
      \left(\frac{\gamma}{\sqrt{2 \pi}}\right)^3 \nonumber \\
  \langle \delij{1}{3} \rangle = \langle \delij{1}{4} \rangle =\langle
  \delij{2}{3} \rangle = \langle \delij{2}{4} \rangle =
  \frac{1}{2} \left(\frac{\gamma}{\sqrt{2 \pi}}\right)^3  \frac{
    \beta^5}{((\gamma^2+\beta^2)/2)^{5/2}} \nonumber \\
  \langle \rho^2 \rangle =   \langle \nu^2 \rangle = \frac{3}{2 \gamma^2}; \; \;  \langle \lambda^2 \rangle  = \frac{5}{2 \beta^2} \nonumber \\
  \langle \vec{\nabla}_\rho^{\,2} \rangle =   \langle \vec{\nabla}_\nu^{\,2} \rangle =
  \frac{3 \gamma^2}{2}; \; \;   \langle \vec{\nabla}_\lambda^{\,2} \rangle = 
  \frac{5 \beta^2}{2}.
\end{eqnarray}
The spin structure of $ \langle \colthree,  \colthree | V^\sigma(q^4)|
\colthree, \colthree \rangle$ guarantees an attractive potential only for
the spin-0, spin-0 correlation,
\begin{equation}\label{spin}
  \langle \colthree,  \colthree | V^\sigma(q^4)| \colthree, \colthree
  \rangle = \frac{h}{6} \left(\frac{\gamma}{\sqrt{2 \pi}}\right)^3
   \left( \begin{array}{c} 
      -6 \\ 2- \frac{\beta^5}{((\beta^2+\gamma^2)/2)^{5/2}} \\
      2+ \frac{1}{2}  \frac{\beta^5}{((\beta^2+\gamma^2)/2)^{5/2}}
    \end{array} \right) \left\{ \begin{array}{l} 0 \otimes 0 \to 0 \\ 1 \otimes 1 \to 0 \\ 1 \otimes 1 \to 2 \end{array} \right.
\end{equation} 

Compare this with the result of the schematic approximation,
eqn(\ref{Schem}). We see that by suitable choice of $C$, the schematic
approximation can duplicate the $0 \otimes 0$ result, but that unless
there is accidental cancellation it would be unable to describe the $1
\otimes 1 \to 0$ result. The origin of the non-zero value for $1
\otimes 1 \to 0$ is the different expectation values of the delta
function with a $P$-wave and without (see equation\eqref{exp}). We propose that this is a general problem
with the schematic approximation when applied to states with
non-trivial spatial dependence and that by its use one can miss
significant physics.

Returning to our study of the quark potential model we see that $\langle \colthree,  \colthree | V^\sigma(\bar{q})|
\colthree, \colthree \rangle$ is particularly simple. The spatial and
colour dependence is identical for all quarks and factors out
leaving a sum of the quark spins, which in the $0 \otimes 0 \to 0$
channel is zero, hence the anti-quark does not change the hyperfine energy
of the system in the Jaffe-Wilczek correlation.

The expectation value of the Hamiltonian is thus,
\begin{equation}\label{ham}
  \langle JW |H|JW \rangle =\left[ \frac{3 \alpha^2}{4 \mu} + \frac{2
    a}{\alpha^2}\right] + \left[\frac{5 \beta^2}{4 m} + \frac{5  a}{2 \beta^2}\right] +  2 \left[ \frac{3 \gamma^2}{4 m} + \frac{3 a}{\gamma^2}\right] - h  \left(\frac{\gamma}{\sqrt{2 \pi}}\right)^3.
\end{equation}
The minimum is found to be at $\alpha=(8 \mu a/3)^{1/4},\, \beta = (2 m
a)^{1/4}$ and $\gamma$ satisfying $ \frac{3 h}{(2 \pi)^{3/2}} \gamma^5
- \frac{3}{m} \gamma^4 + 12 a=0$. We can set the parameters $m, a, h$ using conventional baryon spectroscopy and along the way demonstrate that this model can perfectly well describe the $\Delta-N$ splitting.

$a$ can be set approximately using the $S$-wave $P$-wave
splitting ($\omega_P$) of roughly $500$MeV for the non-strange baryons, since in
the harmonic oscillator $\omega_P = \sqrt{4a/m}$. This gives $a \sim
\frac{(405 \mathrm{MeV})^4}{ 4 \times 330 \mathrm{MeV}}$, where the reason for this unusual presentation will become clear later. If we
consider one-gluon-exchange to be the origin of the contact
term\cite{dgg, IK}, then $h=\frac{8 \pi}{3} \frac{\alpha_S}{m^2}$, where $\alpha_S
\sim 0.75$ is usual. The light quark mass takes its conventional value
$m=330 \mathrm{MeV}$.

The expectation value of the Hamiltonian for the nucleon between
$S$-wave Gaussian trial wavefunctions (with oscillator parameter $\alpha_\rho$) is then
\begin{equation}\label{nuc}
  2 \left[\frac{3 \alpha_\rho^2}{4 m} + \frac{3
      a}{\alpha_\rho^2}\right] - \frac{1}{3}\sqrt{\frac{2}{\pi}}
  \frac{\alpha_S}{m^2}\alpha_\rho^3,
\end{equation}
which is minimised by $\alpha_\rho = 440 \mathrm{MeV}$, with
colour-spin hyperfine energy $-150 \mathrm{MeV}$. This corresponds to
a $\Delta-N$ splitting of $300 \mathrm{MeV}$, in good agreement with
  data\cite{pdg} \footnote{of course, the parameter $\alpha_S$ has been chosen to give this good agreement}. My approach differs from that usually taken which considers
  the hyperfine term only in first order of perturbation theory. By
  using a variational ansatz we are allowing the hyperfine term to
  modify both the energy and the wavefunction ($\alpha_\rho$ would
  have been $405 \mathrm{MeV}$ had the hyperfine term been neglected - hence the rather unusual form of $a$ presented earlier which makes this obvious).

Returning to the Jaffe-Wilczek state, using these parameter values we
find $\alpha = 370$MeV, $\beta = 340$MeV and $\gamma=520$MeV. Had we
neglected the hyperfine term, $\gamma$ would have been $405$MeV. The
hyperfine term is reducing the mean distance between quarks 1 \& 2 and
quarks 3 \& 4 relative to the \{12\}\{34\} distance, which is exactly
what one demands of a spatial diquark-diquark state and what we need
for the unsymmetrised ansatz to be justified. Specifically with these numbers the diquarks have mean radius $\sim 1/3$ fm and are separated by an average distance $\sim 1$ fm.

With $\gamma=520$MeV, the hyperfine energy is $-510$MeV, much larger than the cost of exciting the $P$-wave ($\tfrac{\beta^2}{m} \sim 350$MeV), so that the
Jaffe-Wilczek state is considerably lighter than one would have
naively expected. Simply adding together quark masses and including a $P$-wave energy of $\sim 350$ MeV would suggest a mass over $2$ GeV. Consider the difference $m(\Theta)-m(N)$ which in this model (where the Hamiltonian does not include the quark rest masses) will be
\begin{equation}\label{mtheta}
  \langle JW |H| JW \rangle + m +m_s - \langle N| H |N \rangle .\nonumber
\end{equation}
This is found to be (with $m_s = 450$MeV), $\sim 590$MeV and hence
$m(\Theta) \sim 1530$MeV. This agreement is fortuitous - we have, for
example, neglected tensor interactions which are non-zero due to the
$P$-wave character and which will change this prediction. 

We have observed something rather interesting - A Jaffe-Wilczek-like
$\Theta$ state of low mass with compact diquarks has emerged from a
model which also gives an excellent description of the $\Delta-N$
splitting. From conventional quark model analysis one would not have expected this - the usual
colour-spin arguments as applied to the nucleon suggest a spin-zero ``diquark''
of mass $\sim 600$ MeV which if used naively in the Jaffe-Wilczek scheme would
hugely overpredict the $\Theta$ mass. What we have demonstrated here
is that the diquarks in the $\Theta$ are not like the ``diquark'' in
the nucleon. The quarks in the nucleon all overlap spatially and
correct antisymmetrisation must be carried out. This does not allow
for the kind of spatially distinct diquark found for the $\Theta$,
instead the correlation can only be in spin and flavour. 

\section{Flavour-Spin $\mathbf{210}$ ``ground state''}

The Jaffe-Wilczek correlation would not be our first guess for the ground state of the $q^4 \bar{q}$ system. Without colour-spin
interactions a state with $S$-waves between all quarks would be
expected to have lower energy. Jaffe \& Wilczek suggest that the
colour-spin interactions would force such a state to a higher energy
than their correlation and it is this to which we now turn in this simple
model.

The totally antisymmetric $q^4$ state with symmetric spatial part is
in a $\mathbf{210}$ of flavour-spin and is coupled to total spin
$S(q^4)=1$ \cite{bijker}. It has explicit form
\begin{equation}\label{210}
  \frac{1}{\sqrt{3}}\left\{ |\FSsixone, \FSsixone \rangle \otimes |\colthree, \colthree \rangle  +\left( \frac{\sqrt{3}}{2} |\FSthreeone, \FSthreezero\rangle - \frac{1}{2}|\FSsixzero, \FSsixone \rangle \right) \otimes |\colsix, \colthree \rangle + \left( \frac{\sqrt{3}}{2} |\FSthreezero, \FSthreeone \rangle - \frac{1}{2}|\FSsixone, \FSsixzero \rangle \right) \otimes |\colthree, \colsix \rangle           \right\}_{\bar{\mathbf{6}}, 1 \otimes \mathbf{3}_c},
\end{equation} 
where we use the ``diquark'' notation without implying that diquarks
are dynamically generated. The spatial form (for $q^4$) is
\begin{equation}\label{210space}
  \psi =  N  [Y_{00}(\hat \lambda)e^{- \bar{\beta}^2 \lambda^2 /2}][ Y_{00}(\hat \rho)e^{- \bar{\beta}^2 \rho^2 /2}][
  Y_{00}(\hat \nu) e^{- \bar{\beta}^2 \nu^2 /2}],
\end{equation} 
whose overall symmetry is exposed by expressing the exponent in terms
of the $\vec{r}_i$ as  $- 2 \bar{\beta}^2 \sum_{i>j} (\vec{r}_i -
\vec{r}_j)^2$.

In this flavour-spin-colour state the potential $V(q^4)$ is $\frac{4
  a}{3}(\lambda^2 + \rho^2 + \nu^2)$. A straightforward but somewhat lengthy computation yields the expectation of $V^\sigma(q^4)$ in this state. This is
simplified slightly by the expectation of the delta function, $\langle
\delij{i}{j} \rangle =  \left(\frac{\bar{\beta}}{\sqrt{2
      \pi}}\right)^3$, being $i,j$ independent. We find
\begin{equation}\label{210cs}
  \langle V^\sigma(q^4)\rangle = \frac{8 \pi}{9} \frac{\alpha_S}{m^2} \left(\frac{\bar{\beta}}{\sqrt{2 \pi}}\right)^3,
\end{equation}
which is repulsive, as anticipated. Comparing with the $q^4$ part of
the Hamiltonian in the Jaffe-Wilczek case we find that the
$\mathbf{210}_{FS}$ is around $60$MeV heavier. We have not considered the
colour-spin interaction between $q^4$ and the anti-quark which may be
non-zero due to the $S(q^4)=1$ character of the $\mathbf{210}_{FS}$
state. This could raise the mass of the $\mathbf{210}_{FS}$ even further.

Thus in this simple model with colour-spin contact
interactions a Jaffe-Wilczek-like state can be consistently defined
which is lighter than the naive ground state. The diquark size is small
on the scale of their separation, which was assumed in the spin-orbit
analysis of \cite{spinorbit}.

\section{Smearing the delta function} 

One should worry about the validity of using a delta function
interaction in a Hamiltonian. Such an object is too singular at the
origin to use the normal boundary condition $u(0)=0$ to quantise the
energy levels; we have ignored this problem by using a variational ansatz which satisfies the usual boundary condition. We still have a problem if we consider the origin of this term - it came at
order $(v/c)^2$ in a non-relativistic reduction of the
one-gluon-exchange process, hence in using this term we are implicitly
assuming that the momenta of the quarks are much lower than their
mass. At very small interquark separations the corresponding momentum
scale is rather large and the non-relativistic approximation breaks down. As such we
should ``smear out'' the delta function on the scale of the quark
mass. A suitable modification is
\begin{equation}\label{smear}
  \delta(\vec{r}) \to \frac{1}{(r_0 \sqrt{\pi})^3} \exp [-r^2/r_0^2]
\end{equation}   
with $r_0 \sim m_q^{-1}$. This type of smearing has been used
previously in the literature when modeling conventional hadron
spectra\cite{Brau, GI}. The following smearing scales were found to
be phenomenologically satisfactory:

\begin{tabular}{cc}
\cite{Brau} & $r_0 \sim 1/(1280 \mathrm{MeV}) \sim 1/(4 m_u)$;\\
\cite{GI} & $r_0 \sim 1/(1870 \mathrm{MeV}) \sim 1/(6 m_u)$.
\end{tabular}

We will consider here the effect of smearing on the results we reported
in the previous section. The change is $\langle \beta
|\delta(\vec{r})| \beta \rangle \equiv (\beta/\sqrt{\pi})^3 \to
(\beta/\sqrt{\pi})^3 (r_0^2 \beta^2 +1)^{-3/2}$. With $r_0=1/(n m_u)$,
the following is obtained for the nucleon, Jaffe-Wilczek state and the
$\mathbf{210}_{FS}$

\begin{tabular}{l |cc | cc |cc |}
 & \multicolumn{2}{c}{nucleon} & \multicolumn{2}{c}{JW} &\multicolumn{2}{c}{$\mathbf{210}_{FS}$}\\
  $n$ & $\alpha_\rho$ & $E_{\mathrm{hyp}}$ &$\gamma$ & $E_{\mathrm{hyp}}$&
   $\bar{\beta}$ & $E_{\mathrm{hyp}}$\\
\hline
1 & 408 & -31 & 410 & -62 & 365 &19\\
4 & 428 & -124 & 464 & -307 & 358 & 54\\
6 & 433 & -139 & 483 & -379 & 358 & 57\\
10 & 436 & -147 & 501 & -444 &  358 & 59
\end{tabular}

The delta-function is recovered in the limit $n \to \infty$. So while the nucleon and $\mathbf{210}_{FS}$ state are not strongly affected by the smearing for the phenomenological values $n \sim 4 \to 6$, the large effects felt by the Jaffe-Wilczek
state are diluted considerably. Since the $P$-wave excitation energy is independent of $n$, the dilution is such that 
the Jaffe-Wilczek state is heavier than the $\mathbf{210}_{FS}$.

That the conclusions arrived at in the previous sections are cut-off
dependent is disappointing, but at least the trends remain. The
Jaffe-Wilczek-like state still undergoes a considerable downward shift and
spatially localised diquarks still seem to be formed.

\section{Extensions}

As mentioned in the introduction, an attractive feature of diquark
correlations is their supposed ability to describe the light scalar
mesons. A simple extension to this work would consider
$qq\bar{q}\bar{q}$ states; however, without a $P$-wave to separate the
diquark from the anti-diquark it seems unlikely that spatially
distinct diquarks of the type found earlier will emerge. Of course
such a study would share many similarities with the classic analysis
(in the bag model) of Jaffe\cite{jaffescalar}.

Another interesting state to consider is the $ududud$ state with two
$P$-waves. This has the same quantum numbers as the deuteron. If three
diquarks are formed with $P$-waves between them we would hope that it
has mass greater than the measured mass of the deuteron which we know
to be well described as a bound-state of a proton and a neutron with
small overlap. Consideration of such a state has been advocated in
\cite{spinorbit} as a test on any model proposed to give a light
$\Theta$.

This model, extended to allow unequal quark masses, could predict the
masses of the other members of the pentaquark $\overline{\mathbf{10}}
\oplus \mathbf{8}$, being an explicit realisation of $SU(3)_F$
breaking. In particular this would test the phenomenological
Jaffe-Wilczek Hamiltonian $H_s = M_0 + (n_s + n_{\bar{s}}) m_s + n_s
\alpha$ and its predictions for the other pentaquark states.

If confirmed, the anti-charmed pentaquark observed by H1\cite{H1} with a mass of
$3.1$ GeV would have its mass significantly underpredicted
by this model, as it is in the original Jaffe-Wilczek paper\cite{jw}. Setting the charm quark mass using the experimental
$\Lambda_c - \Lambda$ mass difference and computing the new Hamiltonian expectation, we find a mass of around
$2.8$ GeV for the equivalent anti-charmed Jaffe-Wilczek-like state, which is within $100$ MeV of the Jaffe-Wilczek prediction and very close to the $DN$ threshold. If
the magnitude of spin-orbit splitting calculated in \cite{spinorbit} is correct, the
possibility that the observed state is the $3/2^+$ state is unlikely. One possibility would be that the lightest
anti-charmed pentaquark pair ($1/2^+, 3/2^+$) is still to be found and that the H1 observation is an excited
state, such as the vector diquark excitation. The state $1 \otimes 1
\to 0$ in the model presented in this paper is about $400$MeV heavier than the
Jaffe-Wilczek-like state. This would be a little heavy for the H1
candidate but probably within model errors; alternatively the state with one vector diquark and one scalar
diquark will be somewhat lighter and might be a possibility, but only if the H1 state is the $I_z=0$ part of an isovector.

{\bf Acknowledgements} 
 
The author thanks PPARC for the studentship which funded this
work. Thanks to F.E. Close for inspiration and careful reading of the
manuscript.

\end{document}